\documentclass[runningheads]{llncs}
\usepackage[T1]{fontenc}
\usepackage{graphicx}
\usepackage{bm}
\usepackage{amsmath} 
\usepackage{amsfonts} 
\usepackage{makecell}

\usepackage[hidelinks,colorlinks=true,linkcolor=blue,citecolor=blue,urlcolor=blue]{hyperref}
\begin{document}
\title{Non-rigid Medical Image Registration using Physics-informed Neural Networks}
%
%

 \author{Zhe Min \inst{1} \and
  Zachary M. C. Baum \inst{1} \and
  Shaheer U. Saeed \inst{1}
  \and
  Mark Emberton \inst{2}
\and \\
Dean C. Barratt \inst{1}\and 
 Zeike A. Taylor  \inst{3}
 \and
 Yipeng Hu \inst{1}
 }
\institute{Centre for Medical Image Computing and Wellcome/EPSRC Centre for
Interventional \& Surgical Sciences,  University College London, London, UK
\email{\{z.min,zachary.baum.19,shaheer.saeed.17,d.barratt,yipeng.hu\}@ucl.ac.uk}
\and 
Division of Surgery \& Interventional Science, University College London, London,
UK
\email{m.emberton@ucl.ac.uk}
\and
CISTIB Centre for Computational Imaging and Simulation Technologies in
Biomedicine, Institute of Medical and Biological Engineering, University of Leeds,
Leeds, UK
\email{z.taylor@leeds.ac.uk}
}
 
 \authorrunning{F. Author et al.}
%
\maketitle              
\begin{abstract}

Biomechanical modelling of soft tissue provides a non-data-driven method for constraining medical image registration, such that the estimated spatial transformation is considered biophysically plausible. This has not only been adopted in real-world clinical applications, such as the MR-to-ultrasound registration for prostate intervention of interest in this work, but also provides an explainable means of understanding the organ motion and spatial correspondence establishment. This work instantiates the recently-proposed physics-informed neural networks (PINNs) to a 3D linear elastic model for modelling prostate motion commonly encountered during transrectal ultrasound guided procedures. To overcome a widely-recognised challenge in generalising PINNs to different subjects, we propose to use PointNet as the nodal-permutation-invariant feature extractor, together with a registration algorithm that aligns point sets and simultaneously takes into account the PINN-imposed biomechanics. 
Using 77 pairs of MR and ultrasound images from real clinical prostate cancer biopsy, we first demonstrate the efficacy of the proposed registration algorithms in an ``unsupervised'' subject-specific manner for reducing the target registration error (TRE)
compared to that without PINNs especially for patients with large deformations. The improvements  stem from the intended biomechanical characteristics being regularised, e.g., the resulting deformation magnitude in rigid transition zones was effectively modulated to be smaller than that in softer peripheral zones. This is further validated to achieve low registration error values of $1.90\pm0.52$ mm and $1.94\pm0.59$ mm for all and surface nodes, respectively, based on ground-truth computed using finite element methods.
We then extend and validate the PINN-constrained registration network that can generalise to new subjects. The trained network reduced the rigid-to-soft-region ratio of rigid-excluded deformation magnitude from $1.35\pm0.15$, without PINNs, to $0.89\pm0.11$ $(p<0.001)$ on unseen holdout subjects, which also witnessed decreased TREs from $6.96\pm1.90$ mm to $6.12\pm1.95$ mm $(p=0.018)$. The codes are available at \url{https://github.com/ZheMin-1992/Registration_PINNs}.

\end{abstract}
\section{Introduction}
\indent Multi-modal image registration enables access to clinically important information from different imaging modalities by spatially aligning them \cite{haskins2020deep}, in tasks such as surgical and interventional guidance \cite{hu2012mr,baum2021real}.
Perhaps due to the complementary nature between cross-modality images, designing a robust objective function or an unsupervised loss function is in general highly challenging, for classical or learning-based algorithms, respectively. This work investigates an example of such cross-modality registration, for establishing spatial correspondence between preoperative MR and 3D intraoperative transrectal ultrasound (TRUS) images from the same patients. Indeed, most previously proposed approaches utilised correspondent features from both images, for either iterative optimisation algorithms \cite{van2015biomechanical} 
or neural network training~\cite{hu2018weakly,zeng2020label}. The inevitable sparsity of these available anatomical features, such as the boundaries of prostate gland and other zonal structures, necessitates the addition of transformation smoothness constraints. Hu et al \cite{hu2018weakly} illustrated examples showing that, without imposing smoothness constraints on the registration-estimated transformation, highly distorted local deformation occurred which led to poorer target registration errors (TREs) in these areas. In addition to heuristically designed deformation regularisation, such as L$^2$ norm of local displacement and bending energy, displacement constraints originated from solid mechanics \cite{van2015biomechanical,TowardsPersonalizedStatisticalDeformableModel},
have also demonstrated benefits in this application, with an arguably flexible and purposive approach through its soft tissue modelling physics.



Different from voxelised volumetric images with rectangular grids, point sets are in general unstructured and unordered \cite{qi2017pointnet} 
for efficiently yet sparsely representing geometries or shapes. PointNet was proposed to represent such point sets \cite{qi2017pointnet}. 
Originally designed for classification and segmentation tasks, PointNet was also adopted for learning-based rigid registration that either \textit{1)} first establishes point correspondences in the feature spaces, with which then estimates the rigid transformation using closed-form solutions such as singular value decomposition 
\cite{yew2020rpm}, or \textit{2)} directly aligns with learned feature representations to regress the rigid transformation parameters \cite{Li_2021_CVPR}.
Among non-rigid registration approaches, Free Point Transformer \cite{baum2020multimodality,baum2021real} is an example that utilises the PointNet to extract features to predict source-point-wise displacement vectors, trained with composition of Chamfer loss \cite{fan2017point} and/or negative log-likelihood function of Gaussian Mixture Models \cite{baum2021real}.

 In \cite{saeed2020prostate}, an adapted PointNet \cite{qi2017pointnet} was proposed using finite element modelling (FEM)-simulated training data to predict nodal displacement vectors for prostate meshes with unseen patients. In \cite{fu2021biomechanically}, FEM was first proposed to generate displacements for source point sets with boundary conditions established from an independent non-rigid iterative closest point (ICP) \cite{besl1992method} procedure between prostate surfaces, before a network trained using the FEM-generated transformations \cite{fu2021biomechanically}. Biomechanical constraints have also been investigated in motion modelling and deformable registration, for other organs, such as liver \cite{pfeiffer2020non}, brain \cite{luo2022dataset} and heart \cite{QIN2023102682}. 

This work investigates an alternative approach to encode biomechanical constraints represented by a system of partial differential equations (PDEs), which is solved simultaneously with minimising a registration loss. For registering MR and TRUS prostate images, we propose an approach that \textit{1)} represents prostate point displacements using PointNet, previously adopted in this application \cite{baum2021real}; \textit{2)} develops physics-informed neural network (PINNs) for imposing elastic constraints on the estimated displacements; and \textit{3)} formulates an end-to-end registration network training algorithm, by minimising surface distance as estimated boundary conditions in the PDEs. First, we show that the proposed PINNs effectively constrained the registration-estimated deformation with predefined elastic material properties, for registering individual point pairs. Second, with training data from as few as 75 subjects, the learned constrained registration generalised to new subjects, from which different point sets are independently sampled to represent varying sizes and geometries. We argue in this paper the significance in both results. The subject-specific algorithm incorporates elasticity or potentially other complex constraints in registration in a single network training, replacing alternative biomechnically-constrained methods requiring construction of statistical motion models \cite{hu2011modelling} or finite element simulations \cite{hu2012mr,fu2021biomechanically}; whilst the second learning approach registers unseen point set pairs during efficient inference, demonstrating the generalisability over different geometries and nodal configurations - a well-recognised challenge associated with PINNs.

The contributions are summarised as follows. 
\textbf{1)} 
We developed a patient-specific registration algorithm combining PointNet and PINNs, which aligns prostate glands segmented from MR and TRUS images, subject to biomechanical constraints exerted from soft-tissue-modelling PDEs (Fig. \ref{Illustration_method}).
\textbf{2)} 
We demonstrated that both the biomechanically-regularised deformation and the TRE-reducing correspondence can be generalised to unseen new patients, with the PINN-based registration network trained on a small number of training examples.
\textbf{3)} 
We presented a set of experimental results for evaluating the theoretical and clinical efficacy in soft tissue modelling within registration algorithms, with statistical significance, using finite element (FE)-based ground-truth and independent landmark-based target registration errors (TREs), respectively.

\section{Methods}
\indent 

Let $\mathbf{P}_\mathcal{S}\in\mathbb{R}^{N_s\times3}$ and $\mathbf{P}_\mathcal{T}\in\mathbb{R}^{N_t\times3}$ be a pair of source and target point sets 
with individual points being $\mathbf{p}_s\in\mathbb{R}^{3}$ and $\mathbf{p}_t\in\mathbb{R}^{3}$, where $N_s\in\mathbb{N}^+$ and $N_t\in\mathbb{N}^+$ are number of points, $s\in \{1,...,N_s\}$ and $t\in \{1,...,N_t\}$ are indexes of points. The non-rigid point set registration problem is to find point-wise displacement vectors $\mathbf{D}_\mathcal{S}\in\mathbb{R}^{N_s\times3}$ with $\mathbf{d}_s\in\mathbb{R}^3$, such that the warped source point set 
$\mathsf{T}(\mathbf{P}_\mathcal{S}) = \mathbf{P}_\mathcal{S} + \mathbf{D}_\mathcal{S} $ aligns with $\mathbf{P}_\mathcal{T}$. We additionally adopt notations 
$\mathbf{P}_\mathcal{S}^{\text{internal}}$ and $\mathbf{P}_\mathcal{S}^{\text{surface}}$
to distinguish internal and surface points in $\mathbf{P}_\mathcal{S}$.


\subsection{Physics-informed Neural Network (PINNs) for Non-rigid Registration with Biomechanical Constraints}
\label{PINNs for non-rigid image registration}
\indent With the capability of universal function approximation, physics-informed neural networks (PINNs) can be utilised to model physical laws represented by nonlinear partial differential equations (PDEs) \cite{raissi2019physics}. 
A non-rigid medical image registration problem estimating displacement vectors $\mathbf{D}_\mathcal{S}$ 
is considered as the problem of seeking data-driven solutions to PDEs.
The entire network $e_{\theta}(\mathcal{D}_k)$ where $k \in \mathbb{N}^{+}$ is the patient index, with trainable parameters $\theta$, consists of two sub-networks $g_{\theta_g}(\mathcal{D}_k  
)$ and $h_{\theta_h}(\mathcal{D}_k)$, with completing parameter sets $\theta_g$ and $\theta_h$, predicting displacement vectors $\mathbf{D}_\mathcal{S}$ 
and stress tensors $\bm{\sigma}\in\mathbb{R}^{N_s\times6}$, 
respectively.
Let a function $f(\mathbf{p}_s,\mathbf{d}_s,\bm{\sigma}^s)$ be a PINN defining biomechanical constraints partially characterised by known material properties $b_s$:
\begin{equation}
\label{the overall physics informed neural network}
\begin{tiny} 
  f:= 
  f_1(\frac{\partial \bm{\sigma}^s }{\partial x},\frac{\partial \bm{\sigma}^s }{\partial y}, \frac{\partial \bm{\sigma}^s }{\partial z}) + f_2(\frac{\partial \mathbf{d}_s }{\partial x},\frac{\partial \mathbf{d}_s }{\partial y}, \frac{\partial \mathbf{d}_s }{\partial z}, \bm{\sigma}^s, b_s) + f_3(\frac{\partial \mathbf{d}_s }{\partial x},\frac{\partial \mathbf{d}_s }{\partial y}, \frac{\partial \mathbf{d}_s}{\partial z}, \bm{\sigma}^s),
  \end{tiny}
\end{equation}
where $x$, $y$ and $z$ are spatial coordinates of $\mathbf{p}_s$, the determination of $b_s$ is detailedly described in Sect. \ref{Experiments and Results}, 
$f_1(\cdot)$, $f_2(\cdot)$ and $f_3(\cdot)$ represent norms of residuals deviating from static equilibrium, constitutive equality and null elastic energy, respectively, as defined in the remainder Sect. \ref{the physical laws to be satisfied} and Sect. \ref{non-rigid point set registration using PINNs}.
The network parameters are optimised by minimising $\mathcal{L}^k(\theta;\mathcal{D}_k)=\mathcal{L}_{R}^k(\theta_g; \mathcal{D}_k) +\mathcal{L}_{F}^k(\theta;\mathcal{D}_k)$, where $\mathcal{L}_{F}^k(\theta;\mathcal{D}_k) = \sum_{s=1}^{N_s}  f(\mathbf{p}_s,\mathbf{d}_s,\bm{\sigma}^s)$ is the term concerning biomechanical constraints over all sampled source points, while $\mathcal{L}_{R}^k(\theta_g; \mathcal{D}_k)$ can be either 
(1) $ \sum_{s=1}^{N_s} \big(||\mathbf{d}_s-\mathbf{d}_s^{\text{gt}}||_2^2\big)$ with $\mathbf{d}_s^{\text{gt}}\in\mathbb{R}^3$ denoting 
 ground-truth displacement vectors
  of $\mathbf{p}_s$ under supervised learning (e.g., simulated data with known ground-truth deformations); or (2)   $\phi(\mathsf{T}(\mathbf{P}_\mathcal{S}), \mathbf{P}_\mathcal{T}) $
 being the unsupervised loss (e.g., the Chamfer loss for the purpose of aligning point sets) which measures goodness-of-prediction, resulting in a complete registration algorithm as described in Sect. \ref{non-rigid point set registration using PINNs} and used throughout this paper.


\subsection{Governing Equations for Deforming Linear Elastic Organs adapted for Medical Image Registration}
\label{the physical laws to be satisfied}
\indent 
In this section, linear elasticity is used as a specific example of prostate gland deformation between $\mathbf{P}_\mathcal{S}$ 
and $\mathbf{P}_\mathcal{T}$, primarily due to contact with a moving ultrasound probe \cite{hu2011modelling,hu2012mr}. Adopting linear elasticity aims to demonstrate the feasibility of modelling soft tissue with the PDE-representing physics as the first step towards more complex and potentially more realistic models, such as nonlinear strain, alternative stress, time-dependent viscoelasticity and plasticity.\\
\noindent \textbf{ Strain-displacement Equations}
 The strain-displacement equation (i.e., kinematic equation) at a source point $\mathbf{p}_s$ is
\begin{equation}
\label{the_overall_strain_displacement_equation}
    \bm{\varepsilon}^s = \frac{1}{2}( \nabla \mathbf{d}_s + \nabla \mathbf{d}_s^{\mathsf{T}}),
\end{equation}
where  $\bm{\varepsilon}^s$ is the infinitesimal second-order Cauchy strain tensor at $\mathbf{p}_s$, $\nabla \mathbf{d}_s$ is the 
displacement gradient w.r.t. spatial coordinates $x,y,z$ of $\mathbf{p}_s$. Eq. (\ref{the_overall_strain_displacement_equation}) can be rewritten explicitly as 
$\varepsilon_{xx}^s = \frac{\partial d_s^x }{\partial x }$,
$\varepsilon_{xy}^s = \frac{1}{2}(\frac{\partial d_s^x }{\partial y}+\frac{\partial d_s^y}{\partial x})$,
$\varepsilon_{yy}^s = \frac{\partial d_s^y }{\partial y}$, 
$\varepsilon_{yz}^s = \frac{1}{2}(\frac{\partial d_s^y }{\partial z}+\frac{\partial d_s^z }{\partial y})$,
$\varepsilon_{zz}^s = \frac{\partial d_s^z}{\partial z} $,
$\varepsilon_{xz}^s = \frac{1}{2}(\frac{\partial  d_s^x}{\partial z }+\frac{\partial  d_s^z}{\partial x })$. Eq. (\ref{the_overall_strain_displacement_equation}) is used to compute strain tensors  $\mathcal{E}\in\mathbb{R}^{N_s\times6}$   
from displacement vectors $\mathbf{D}_\mathcal{S}$
 predicted by $g_{\theta_g}(\mathcal{D}_k  
)$. 
\\
\noindent \textbf{Static Equilibrium Equations}
The spatial components of the Cauchy stress tensor $\bm{\sigma}^s$ at $\mathbf{p}_s$, predicted by $h_{\theta_h}(\mathcal{D}_k)$,   satisfy the following equilibrium equation (i.e., equation of motion)
\begin{equation}
    \label{static equilibrium equation}
    \sigma_{ji,j}^s+ F_{i}=0,
\end{equation}
where 
$(\cdot)_{,j}^s$ is a shorthand for $\frac{\partial{(\cdot)}}{
\partial(\mathbf{p}_s)_j
}$
 , $F_i\in\mathbb{R}$ is the body force that is approximated to be zero at the static equilibrium, $i$ and $j$ denote three spatial directions. Eq. (\ref{static equilibrium equation})  can be rewritten explicitly as
$\frac{\partial \sigma_{xx}^s }{\partial x}
+\frac{\partial \sigma_{yx}^s }{\partial y}
+\frac{\partial \sigma_{zx}^s}{\partial z}
= 0$, $\frac{\partial \sigma_{xy}^s}{\partial x}
    +\frac{\partial \sigma_{yy}^s}{\partial y}
    +\frac{\partial \sigma_{zy}^s}{\partial z}
    = 0$, $\frac{\partial \sigma_{xz}^s}{\partial x}
    +\frac{\partial \sigma_{yz}^s}{\partial y}
    +\frac{\partial \sigma_{zz}^s}{\partial z}
    = 0$.\\
\noindent \textbf{Constitutive Equations}
The stress and strain tensors at $\mathbf{p}_s$ are related by the constitutive equation (i.e., the generalised Hooke's law) as
\begin{equation}
    \label{constitutive equation short form}
    \bm{\sigma}^s = \mathsf{C}:\bm{\varepsilon}^s,
\end{equation}
where $\mathsf{C}$ is the fourth-order elasticity tensor.
Eq. (\ref{constitutive equation short form}) can be expanded as 
\begin{equation}
   \label{constitutive equation long form}
    \begin{split}
    \begin{bmatrix}
        \sigma_{xx}^s\\
        \sigma_{yy}^s\\
        \sigma_{zz}^s\\
        \sigma_{xy}^s\\
        \sigma_{xz}^s\\
        \sigma_{yz}^s
    \end{bmatrix}
    =  
    \begin{bmatrix}
    (\lambda+2\mu) &  \lambda &  \lambda & 0 &0 &0\\
    \lambda     & (\lambda+2\mu)  &  \lambda & 0 &0 &0\\
     \lambda &  \lambda    &(\lambda+2\mu) &  0 &0 &0\\
     0 & 0 &0& \mu &0&0\\
       0 & 0 &0 &0& \mu&0\\
     0 & 0 &0 &0&0& \mu\\
    \end{bmatrix}
    \begin{bmatrix}
        \varepsilon_{xx}^s\\
        \varepsilon_{yy}^s\\
        \varepsilon_{zz}^s\\
        2\varepsilon_{xy}^s\\
        2\varepsilon_{xz}^s\\
        2\varepsilon_{yz}^s
    \end{bmatrix} 
    \end{split},
\end{equation}
where $\lambda \in \mathbb{R}$ and $\mu \in \mathbb{R}$ are Lame parameters, which are computed using $\lambda = \frac{E\nu}{(1-2\nu) (1+\nu)}$ and $\mu =\frac{E}{2(1+\nu)} $ with the Young's Modulus $E$ and Possion's ratio $v$.\\
\indent 
As will be introduced in Sect. \ref{non-rigid point set registration using PINNs}, Eq. (\ref{static equilibrium equation}) and Eq. (\ref{constitutive equation long form}) are utilised to construct PDEs that
regularise $\mathbf{D}_\mathcal{S}$ 
predicted by $g_{\theta_g}(\mathcal{D}_k)$, 
$\bm{\sigma}$ 
predicted by $h_{\theta_h}(\mathcal{D}_k)$, and 
$\mathcal{E}$ 
computed with Eq. (\ref{the_overall_strain_displacement_equation}). 

\begin{figure}[t!]
\includegraphics[width=\textwidth]
{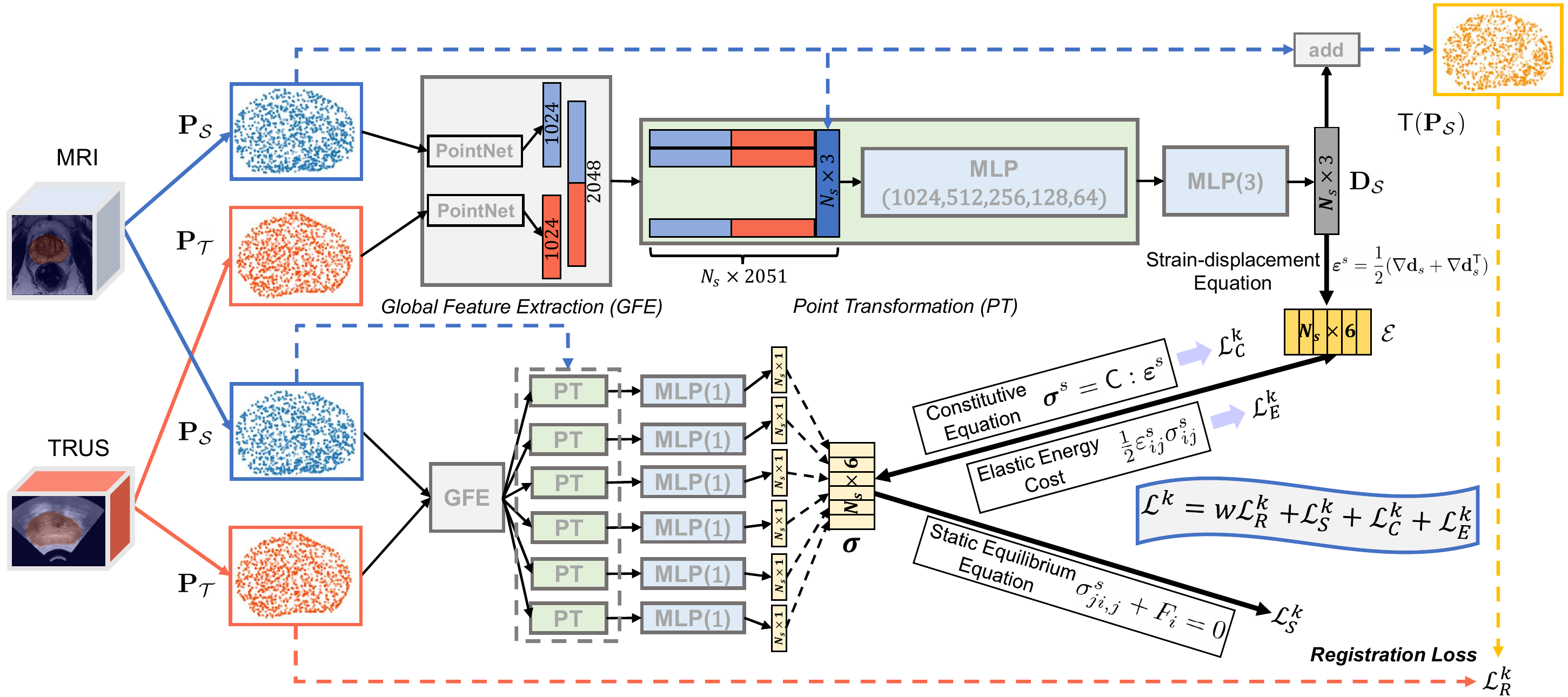}
 \caption{The proposed non-rigid medical image registration framework using physics-informed neural networks (PINNs), whose inputs are a pair of source and target point sets $\mathbf{P}_\mathcal{S}$ and $\mathbf{P}_\mathcal{T}$ extracted from MRI and TRUS volumes of the same patient, respectively. PINNs consist of $g_{\theta_g}(\mathcal{D}_k  
)$ predicting  displacement vectors $\mathbf{D}_\mathcal{S}$ from which the point-wise strain tensors $\mathcal{E}$ 
 are further computed with the strain-displacement equation in Eq. (\ref{the_overall_strain_displacement_equation}), and $h_{\theta_h}(\mathcal{D}_k)$ predicting stress tensors  $\bm{\sigma}$. The source point set $\mathbf{P}_\mathcal{S}$ added by $\mathbf{D}_\mathcal{S}$ results in the warped source point set $\mathsf{T}(\mathbf{P}_\mathcal{S})$, between which and $\mathbf{P}_\mathcal{T}$ the Chamfer loss $\mathcal{L}_{R}^k$  is computed using Eq. (\ref{the equation of LR}). $\mathcal{L}_{S}^k$ in Eq. (\ref{LS}) and $\mathcal{L}_{C}^k$ in Eq. (\ref{LC}) penalise deviations from the equality of the static equilibrium equation in Eq. (\ref{static equilibrium equation}) about $\bm{\sigma}$, and that of the constitutive equation in Eq. (\ref{constitutive equation short form})  about $\bm{\sigma}$ and $\mathcal{E}$, respectively. $\mathcal{L}_{E}^k$ in Eq. (\ref{elastic energy term}) is the elastic energy cost that shall also be minimised.}
\label{Illustration_method}
\end{figure}
\subsection{A Non-Rigid Point Set Registration Algorithm using PINNs}
\label{non-rigid point set registration using PINNs}
\indent  Fig. \ref{Illustration_method} shows the schematic of the proposed non-rigid point set registration network, with the displacement-predicting $g_{\theta_g}(\mathcal{D}_k  
)$ and stress-predicting $h_{\theta_h}(\mathcal{D}_k)$. \\
\noindent 
\textbf{Loss Functions for Single-Pair Patient-specific Registration} 
The loss function includes four terms. First,  the Chamfer loss $\phi(\mathsf{T}(\mathbf{P}_\mathcal{S}), \mathbf{P}_\mathcal{T})$ \cite{fan2017point}
is minimised to spatially align the two point sets, and is given by
\begin{equation}
\label{the equation of LR}
\begin{tiny}
\mathcal{L}_{R}^k(\theta_g;\mathcal{D}_k)= \frac{1}{|\widetilde{N}_t|} \Bigg(  \sum_{t\in \widetilde{N}_t}
        \min_{s\in \widetilde{N}_s} 
        ||\mathsf{T}(\mathbf{p}_s)-  \mathbf{p}_t  ||_2^2
        \Big)
        +         \frac{1}{|\widetilde{N}_s|} \Big(  \sum_{s\in \widetilde{N}_s}
        \min_{t\in \widetilde{N}_t} 
        ||\mathsf{T}(\mathbf{p}_s)-  \mathbf{p}_t  ||_2^2
        \Bigg)
\end{tiny}
\end{equation}
where $\widetilde{N}_s\subseteq   \{1,...,N_s\}$ and $\widetilde{N}_t\subseteq   \{1,...,N_t\}$ denote sets of points being either the entire organ $\mathbf{P}_\mathcal{S}$ 
and $\mathbf{P}_\mathcal{T}$ 
or a subset region, e.g., surface points 
$\mathbf{P}_\mathcal{S}^{\text{surface}}$ and
$\mathbf{P}_\mathcal{T}^{\text{surface}}$,
$|\widetilde{N}_s|$ and $|\widetilde{N}_t|$ are numbers of points. 
Second, deviation from the static equilibrium equation in Eq. (\ref{static equilibrium equation}) w.r.t. the stress $\bm{\sigma}$ is penalised by minimising $\mathcal{L}_{S}^k(\theta_h;\mathcal{D}_k)$ as 
\begin{equation}
\label{LS}
\begin{tiny} 
\mathcal{L}_{S}^k(\theta_h;\mathcal{D}_k)=\sum_{s=1}^{N_s} f_1(\frac{\partial \bm{\sigma}^s }{\partial x},\frac{\partial \bm{\sigma}^s }{\partial y}, \frac{\partial \bm{\sigma}^s }{\partial z}),
\end{tiny}
\end{equation}
where $f_1(\frac{\partial \bm{\sigma}^s }{\partial x},\frac{\partial \bm{\sigma}^s }{\partial y}, \frac{\partial \bm{\sigma}^s }{\partial z}) =
|    \frac{\partial \sigma_{xx}^s}{\partial x}
    +\frac{\partial \sigma_{yx}^s}{\partial y}
    +\frac{\partial \sigma_{zx}^s}{\partial z}| +|
      \frac{\partial \sigma_{xy}^s}{\partial x}
    +\frac{\partial \sigma_{yy}^s}{\partial y}
    +\frac{\partial \sigma_{zy}^s}{\partial z}| 
    + |\frac{\partial \sigma_{xz}^s}{\partial x}
    +\frac{\partial \sigma_{yz}^s}{\partial y}
    +\frac{\partial \sigma_{zz}^s}{\partial z}|$. 
Third, $\mathcal{L}_{C}^k(\theta;\mathcal{D}_k)$ regularises $\bm{\sigma}$ and strain $\mathcal{E}$ 
to satisfy constitutive equations in Eq. (\ref{constitutive equation long form}), and is defined as 
\begin{equation}
\label{LC}
\begin{tiny} 
       \mathcal{L}_{C}^k(\theta;\mathcal{D}_k)= \sum_{s=1}^{N_s} 
       f_2(\frac{\partial \mathbf{d}_s }{\partial x},\frac{\partial \mathbf{d}_s }{\partial y}, \frac{\partial \mathbf{d}_s }{\partial z}, \bm{\sigma}^s, b_s),
\end{tiny}
\end{equation}
where $f_2(\frac{\partial \mathbf{d}_s }{\partial x},\frac{\partial \mathbf{d}_s }{\partial y}, \frac{\partial \mathbf{d}_s }{\partial z}, \bm{\sigma}^s, b_s)= (
    |(\lambda+2\mu)\varepsilon_{xx}^s  +\lambda (\varepsilon_{yy}^s 
    + \varepsilon_{zz}^s) -\sigma_{xx}^s |  
    + |\lambda (\varepsilon_{xx}^s+\varepsilon_{zz}^s) + (\lambda+2\mu) \varepsilon_{yy}^s   -\sigma_{yy}^s | +  | \lambda \varepsilon_{xx}^s + \lambda \varepsilon_{yy}^s  + (\lambda+2\mu) \varepsilon_{zz}^s -\sigma_{zz}^s   | + |\sigma_{xy}^s-2\mu\varepsilon_{xy}^s| + |\sigma_{xz}^s-2\mu\varepsilon_{xz}^s 
    | + |\sigma_{yz}^s-2\mu\varepsilon_{yz}^s | )$, the strain tensor
$\varepsilon_{ij}^s$
 at $\mathbf{p}_s$ is computed from network-predicted $\mathbf{d}_s$ with the automatic differentiation, 
 according to the kinematic equation in Eq. (\ref{the_overall_strain_displacement_equation}).
Fourth, $\mathcal{L}_E^k(\theta;\mathcal{D}_k)  = \sum_{s=1}^{N_s}\frac{1}{2}\varepsilon_{ij}^s\sigma_{ij}^s $ is the elastic energy cost to be minimised
\begin{equation}
\label{elastic energy term}
\begin{tiny} 
   \mathcal{L}_{E}^k(\theta;\mathcal{D}_k) = \sum_{s}^{N_s} 
   f_3(\frac{\partial \mathbf{d}_s }{\partial x},\frac{\partial \mathbf{d}_s }{\partial y}, \frac{\partial \mathbf{d}_s}{\partial z}, \bm{\sigma}^s),
\end{tiny}
\end{equation}
where $f_3(\frac{\partial \mathbf{d}_s }{\partial x},\frac{\partial \mathbf{d}_s }{\partial y}, \frac{\partial \mathbf{d}_s}{\partial z}, \bm{\sigma}^s)=  \frac{1}{2}(\varepsilon_{xx}^s\sigma_{xx}^s
    + \varepsilon_{yy}^s\sigma_{yy}^s
    + \varepsilon_{zz}^s\sigma_{zz}^s 
    + 2\varepsilon_{xy}^s\sigma_{xy}^s
    +2 \varepsilon_{xz}^s\sigma_{xz}^s
    + 2\varepsilon_{yz}^s\sigma_{yz}^s
    )$.\\
\indent The overall training loss $\mathcal{L}^k(\theta;\mathcal{D}_k)$ in the single-pair image registration for the given subject $k$ is given by a  ($w\in\mathbb{R}^{+}$)-weighted sum of these terms, 
\begin{equation}
\label{the overall loss function}
    \mathcal{L}^k(\theta;\mathcal{D}_k) =
    w\mathcal{L}_{R}^k(\theta_g;\mathcal{D}_k)
    + \mathcal{L}_{S}^k(\theta_h,\mathcal{D}_k)
    + \mathcal{L}_{C}^k(\theta;\mathcal{D}_k)
    +\mathcal{L}_{E}^k(\theta;\mathcal{D}_k).
\end{equation}

\noindent \textbf{Optimisation for a Multi-Patient Learning Algorithm}
The above-described network can be adapted with minimal change in implementation, for a population-trained registration algorithm, by optimising network parameters $\theta$ with respect to an amortization loss:
\begin{equation}
    \label{optimisation problem}
    \theta^{\star} = \arg \min_{\theta} \mathcal{L}(\theta;\mathcal{D})=\arg \min_{\theta} E_{k} (\mathcal{L}^k(\theta;\mathcal{D}_k)),
\end{equation}
where $\mathcal{D}$ is all the training data from multiple subjects and $E_k$ denotes the expected value over all training examples.

\begin{figure}[b!]
\includegraphics[width=\textwidth]
{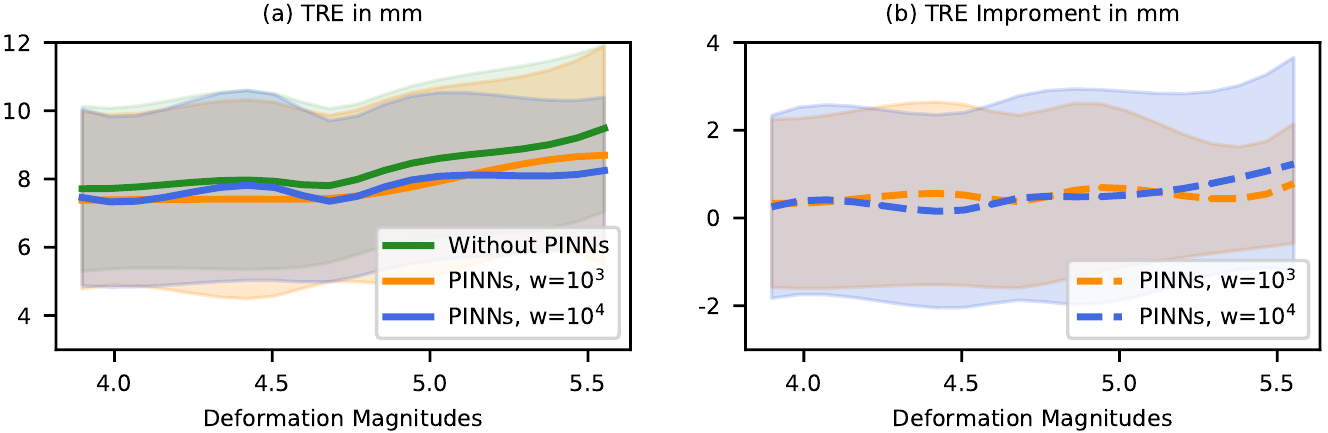}
\caption{ (Left) TRE (mm) of MRI-TRUS fusion using patient-specific registration models with and without PINNs, w.r.t. deformation magnitudes of prostate gland. (Right) TRE improvements by incorporating PINNs compared to those without PINNs. The shaded areas are $95\%$ confidence intervals (i.e., $\pm2$ standard deviations). } 
\label{TRE and Improvments}
\end{figure}
\subsection{Evaluation Metrics} 
For the experimental results described in Sec.~\ref{Experiments and Results}, four evaluation metrics are reported. First, TREs were computed as the average distance between the geometric centroids of pairs of registered source and target landmarks, which include apex and base of the prostate, water-filled cysts, and calcifications. Further details in defining these independent landmarks followed published methods in previous studies \cite{baum2021real,hu2018weakly}. Second, deformation magnitudes (\textsf{DMs}) were computed to measure the ``pure'' non-rigid part of predicted displacements of 
$\mathbf{P}_\mathcal{S}$, 
excluding the ``largest'' rigid transformation $(\mathbf{R}, \mathbf{t})$. \textsf{DM} was defined as residuals after solving the orthogonal Procrustes problem between $\mathbf{P}_\mathcal{S}$ and $\mathsf{T}(\mathbf{P}_\mathcal{S})$
\cite{besl1992method}:  
as $\textsf{DM} = 
   \frac{1}{|\widetilde{N}|}  \sum_{s\in \widetilde{N}}
    || \mathbf{R}\mathbf{p}_s+\mathbf{t} -  \mathsf{T}(\mathbf{p}_s) ||_2$, 
where $\widetilde{N}$ can be either $\mathbf{P}_\mathcal{S}$ or $\mathbf{P}_\mathcal{S}^{\text{internal}}$
, $|\widetilde{N}|$ is the number of points. 
Third, Chamfer Distance (\textsf{CD}) was defined as 
$   \textsf{CD} = \frac{1}{2} \Big(\frac{1}{|\widetilde{N}_t|}  \sum_{t\in \widetilde{N}_t}
        \min_{s \in \widetilde{N}_s} 
        ||\mathsf{T}(\mathbf{p}_s)-  \mathbf{p}_t  ||_2
        +         \frac{1}{|\widetilde{N}_s|}  \sum_{s\in \widetilde{N}_s}
        \min_{t\in \widetilde{N}_t} ||\mathsf{T}(\mathbf{p}_s)-  \mathbf{p}_t  ||_2
         \Big)$,
where $\widetilde{N}_t$ and $\widetilde{N}_s$ are the same as those in Eq. (\ref{the equation of LR}). 
Fourth, the root-mean-square error (\textsf{rmse}) was defined between predicted displacement $\mathbf{D}_\mathcal{S}$
and ground-truth $\mathbf{D}_\mathcal{S}^{gt}\in\mathbb{R}^{|\widetilde{N}_s|\times3}$
as
$   \textsf{rmse} =  \sqrt{ \frac{1}{|\widetilde{N}_s|}\sum_{s\in\widetilde{N}_s} ||\mathbf{d}_s-  \mathbf{d}_s^{\text{gt}}||_2^2  }$.


\section{Experiments and Results}
\label{Experiments and Results}
\noindent \textbf{Datasets}  The first dataset contained 77 pairs of MRI and TRUS volumetric images (both were resampled to 0.8 × 0.8 × 0.8 mm$^3$) from prostate cancer biopsy, where the exemplar clinical application is to register pre-operative MRI images with TRUS images where prostate gland has been deformed due to surgical probe contact \cite{hu2011modelling}. Each pair of point sets was extracted from the segmentations of the prostate gland in one patient's MRI and TRUS images respectively (Fig. \ref{Illustration_method}).
The second dataset containing 8 cases was generated over MRI-derived prostate meshes by producing ground-truth deformations in  $[5.58, 8.66]$ mm using the finite element modelling (FEM) process, proposed in previous studies \cite{hu2011modelling,saeed2020prostate}, with different material properties assigned to peripheral zones (PZ) and transition zone (TZ): the ratios of Youngs' Modulus with PZ and TZ $\frac{E_{\text{PZ}}}{E_{\text{TZ}}}$ were in the range of $[0.12, 0.20]$. More details about zonal segmentations in this dataset can be found in \cite{hu2012mr,hu2011modelling}. 
The third dataset included 75 MRI and TRUS point-set pairs for training and 33 pairs from different patients for testing, in order to validate the generalisability of the developed population-trained model. 
\\
\begin{table}[t] 
\scriptsize
\centering
\caption{Quantitative results (mean $\pm$ std in mm) of patient-specific models in the first experiment. $^{\star}$:  significantly different from results without PINNs ($p<0.05$). $^{\circ}$: the two with $^{\circ}$ had no significant difference with each other but were both significantly different from the remaining result in one column ($p<0.001$). The best results are marked in bold for \textsf{CD} and TRE. }
\label{tab1}
\begin{tabular}{p{62pt} p{56pt} p{56pt} p{52pt} p{56pt} p{39pt}}
\hline
{\makecell[c]{Models}} & {\makecell[c]{ \textsf{DM}\\(Internal Points\\Rigid Region)}}  & {\makecell[c]{ \textsf{DM}\\(Internal Points\\Soft Region)}} 
 & \makecell[c]{\textsf{CD}} & {\makecell[c]{ \textsf{CD}\\(Surface Points\\Only)}} & \makecell[c]{\textsf{TRE}} \\
\hline
{\makecell[c]{  Without PINNs}} 
   & {\makecell[c]{$4.78\pm1.06$}} & {\makecell[c]{$4.72\pm0.95$}} & {\makecell[c]{ $\textbf{1.51}^{\circ}\pm\mathbf{0.23}^{\circ}$}} & {\makecell[c]{ $\textbf{0.49}^{\circ}\pm\mathbf{0.10}^{\circ}$}} &  {\makecell[c]{ $7.52\pm2.46$}} \\ 
 {\makecell[c]{  PINNs ($w=10^4$) } }
&{\makecell[c]{$4.56^{\star}\pm1.14^{\star}$}} & {\makecell[c]{ $4.69\pm0.99$}} & {\makecell[c]{ $1.52^{\circ}\pm0.26^{\circ}$}} & {\makecell[c]{ $0.53^{\circ}\pm0.19^{\circ}$}} & {\makecell[c]{$7.32\pm2.60$}} \\
 {\makecell[c]{  PINNs ($w=10^3$)}} 
 & {\makecell[c]{$4.42^{\star}\pm1.31^{\star}$}} & {\makecell[c]{ $4.58\pm1.06$}} & {\makecell[c]{ $1.66\pm0.30$}} &{\makecell[c]{ $0.83\pm0.42$}}& $\textbf{7.23}\pm\textbf{2.60}$
 \\
\hline
\end{tabular}
\end{table}

\begin{figure}[htbp]
\includegraphics[width=\textwidth]
{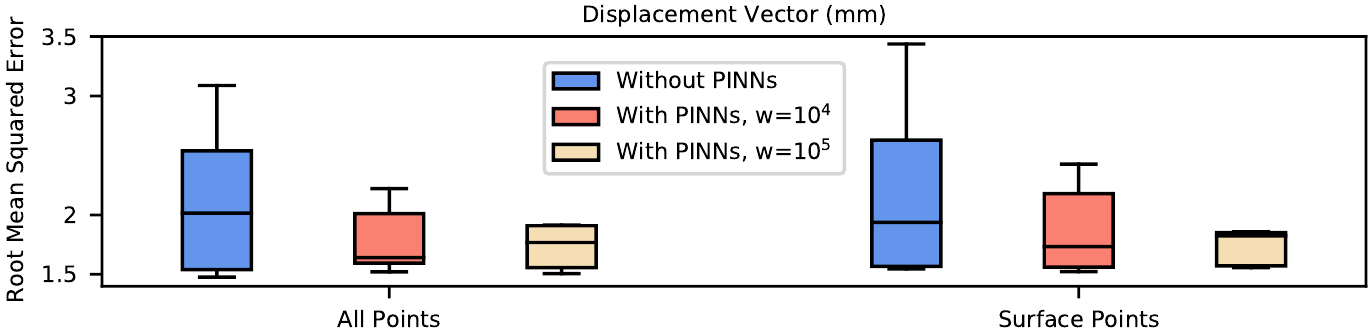}
\caption{ Root mean squared error computed with displacement vectors $\mathbf{D}_\mathcal{S}$ predicted by registration methods and ground-truth $\mathbf{D}_\mathcal{S}^{gt}$ generated with finite element modelling.   } 
\label{Root mean squared error on the simulated data}
\end{figure}

\textbf{Implementation Details} PointNet \cite{qi2017pointnet} is adapted with a TNet 4-by-4 outputting $4\times4$ rigid transformation matrix instead of the original 3-by-3 TNet, suggested in \cite{baum2021real}. 
The final global feature from a PointNet $\phi(\cdot)$ is of size 1024. In the global feature extraction (Fig. \ref{Illustration_method})
module, the global features $\phi(\mathbf{P}_\mathcal{S})$ and  $\phi(\mathbf{P}_\mathcal{T})$ learnt from 
$\mathbf{P}_\mathcal{S}$ and $\mathbf{P}_\mathcal{T}$ are concatenated.
In the point transformation module (Fig. \ref{Illustration_method}), the concatenated global feature is repeated for $N_s$ times and further concatenated with $\mathbf{P}_\mathcal{S}$.  
The resulting feature map of size $N_s\times2051$ will go through shared \textsf{MLP}(1024, 512, 256, 128, 64) and another shared \textsf{MLP}(256) without the ReLU layer. At the end, \textsf{MLP}(3) and  6 individual \textsf{MLP}(1) are used in branches $g_{\theta_g}(\mathcal{D}_k  
)$ predicting 
$\mathbf{D}_\mathcal{S}$ and $h_{\theta_h}(\mathcal{D}_k)$ predicting $\bm{\sigma}$,
respectively. \\
\indent For the first and third experiments, Young's modulus $E$ in Eq. (\ref{LC}) was chosen as $500$ kPa and $5$ kPa for points in rigid and soft compartments while Possion' ratio $v$ was $0.49$, leading to $(\lambda=8221.48, \mu=167.78)$
and $(\lambda=82.21, \mu=1.68)$, respectively. For the second experiment,  $E$ and $v$ were set according to the ratio of their ground-truth values in two sub-regions. 
The two compartments' points were determined either by approximately taking upper $\frac{2}{3}$ and lower $\frac{1}{3}$ sub-regions in the axial view as rigid and soft compartments (as in the first and third experiments with clinical data), or taking the TZ and PZ respectively if zonal segmentations were available (as in the second experiment) \cite{hu2011modelling}. All three experiments were run on an Intel(R) Xeon(R) Gold 5215 CPU with an NVIDIA Quadro GV100 32GB GPU. 
\\
\noindent \textbf{Results} 
Table \ref{tab1} and Fig. \ref{TRE and Improvments} include the numerical results of the first experiment. Two observations can be made from Table \ref{tab1}: \textit{1)} TRE values decreased with PINNs; and more importantly \textit{2)} \textsf{DM} values in the rigid sub-regions were smaller than those in the soft sub-regions with PINNs, which demonstrated biomechanical constraints are effectively preserved in the registration algorithm, i.e., $\frac{\textsf{DM}_{\text{rigid}}} {\textsf{DM}_{\text{soft}}}<1$, significantly different from $\frac{\textsf{DM}_{\text{rigid}}} {\textsf{DM}_{\text{soft}}}>1$ without PINNs ($p=0.019$ for $w=10^3$ and $p=0.029$ for $w=10^4$, paired t-tests at significance level $\alpha$=0.05).  Fig. \ref{TRE and Improvments} shows TRE values w.r.t. varying \textsf{DM} thresholds. 
It is found from Fig. \ref{TRE and Improvments} that \textit{1)} TRE values increased with larger deformation magnitudes for all methods; and \textit{2)} PINNs reduced the TREs and demonstrated greater improvements for patients that undergo larger deformations. For example, PINNs ($w=10^3$) significantly decreased TRE from $7.87\pm2.03$ mm without PINNs to $7.12\pm2.21$ ($p=0.049$), among top $40\%$ (31/77) patients with larger deformations. 
\\
 \begin{table}[t!]
 \scriptsize
\centering
\caption{TRE statistics (mean $\pm$ std in mm) of patients whose ratios of \textsf{DM} between rigid and soft sub-regions were correctly modulated from $>1$ without PINNs to $<1$ with PINNs. 
(1) The $1^{st}$ and $3^{rd}$ rows are such cases; and (2) the $2^{nd}$ and $4^{th}$ rows are patients in (1) whose TRE were also improved. The column ${\#}$ records the number of patients per row. $^{\star}$: improvements of PINNs were statistically significant ($p<0.001$).
}
\label{TRE values for those patients where DM magnitudes have been corrected}
\begin{tabular}{p{100pt} p{20pt} p{65pt} p{60pt} p{65pt}}
\hline
{\makecell[c]{PINNs Models}} &  {\makecell[c]{${\#}$}}  & {\makecell[c]{TRE \\ Without PINNs}} &  {\makecell[c]{TRE \\ With PINNs}} &  {\makecell[c]{TRE Improved\\With PINNs}} \\
\hline
 \makecell[c]{ $w=10^4$}& {\makecell[c]{22}}  & \makecell[c]{$7.39\pm2.08$} & {\makecell[c]{$\textbf{6.55}\pm\textbf{1.86}$}} 
 & {\makecell[c]{$0.84\pm2.18$}} \\
 \makecell[c]{$w=10^4$
 (TRE Improved)}& {\makecell[c]{15}}  & \makecell[c]{$8.04\pm2.05$} & {\makecell[c]{$\textbf{6.04}^{\star}\pm\textbf{1.70}^{\star}$}}& {\makecell[c]{$2.00\pm1.49$}} \\ 
 \makecell[c]{$w=10^3$}& {\makecell[c]{20}} & \makecell[c]{$7.95\pm2.15$} & {\makecell[c]{$\textbf{6.68}^\star\pm\textbf{2.23}^\star$}}   & {\makecell[c]{$1.27\pm1.90$}} \\
  \makecell[c]{$w=10^3$
  (TRE Improved)}&{\makecell[c]{16}}  & \makecell[c]{$8.09\pm2.31$} & {\makecell[c]{$\textbf{6.21}^{\star}\pm\textbf{2.15}^{\star}$}}  & {\makecell[c]{$1.88\pm1.62$}}  \\
\hline
\end{tabular}
\end{table}
\begin{figure}[t!]
\includegraphics[width=\textwidth]{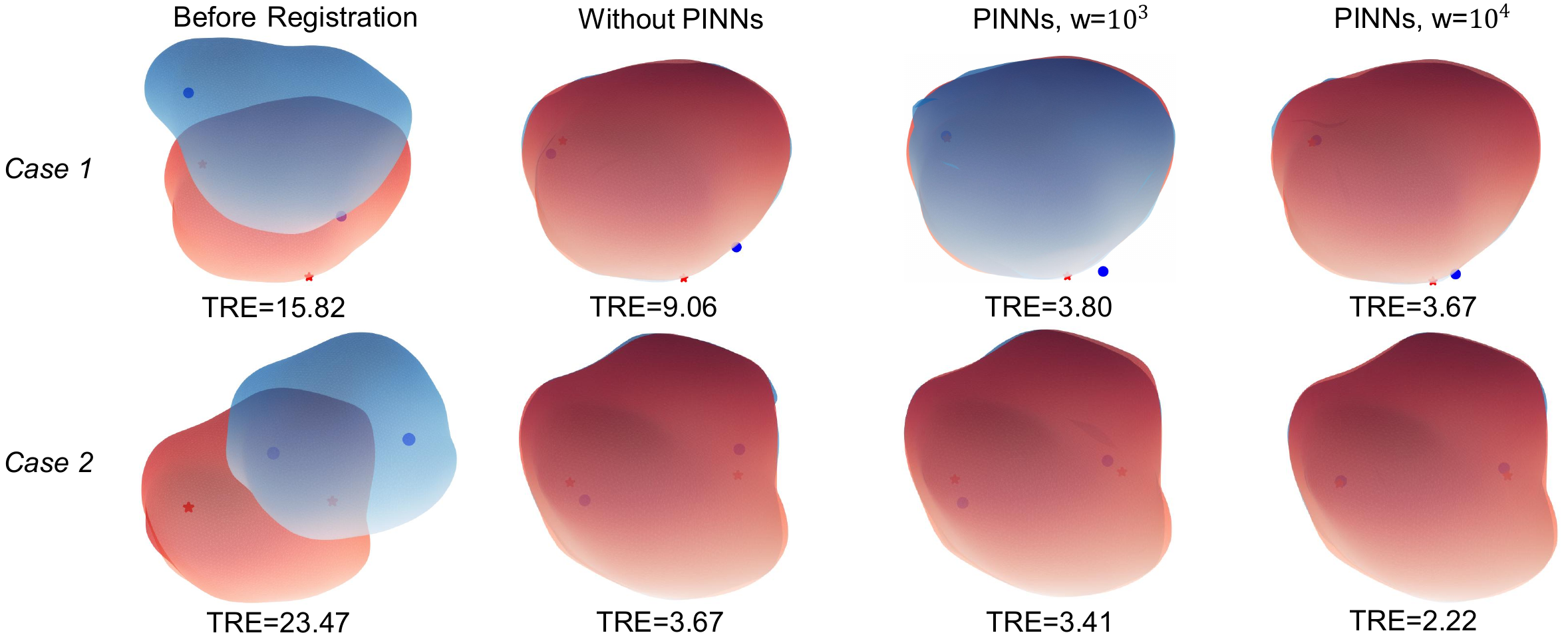}
\caption{Qualitative results showing meshes and TREs (mm) before and after registration, of two patient cases with large deformations. The original and warped source (i.e., MRI) meshes
are depicted in blue while target (i.e., TRUS) meshes are in red. The anatomical landmarks in MRI and TRUS are denoted by blue circles and red 
stars.
} 
\label{qualitative results of patient 49}
\end{figure}
\indent Fig. \ref{qualitative results of patient 49} shows qualitative results from two patients, with large and moderate-to-large non-rigid deformations being 6.30 mm and 5.56 mm. 
Take \textit{case 1} as an example, the registration method with PINNs reached desired smaller \textsf{DM} value in the rigid compartment than that in the soft one, being 3.23 mm versus 3.82 mm for PINNs ($w=10^4$) and 3.25 mm versus 4.04 mm for PINNs ($w=10^3$), whereas without PINNs DM was larger (i.e., 5.26 mm) in the rigid compartment than that (i.e., 4.54 mm) in the soft one. While surface points are visually well aligned for both methods (Fig. \ref{qualitative results of patient 49}) with Chamfer distances 0.83 mm, 0.45 mm and 0.48 mm for PINNs ($w=10^4$), PINNs ($w=10^3$) and without PINNs, PINNs greatly reduced the TRE value from that 
without PINNs (i.e., 
from 9.06 mm to 3.80 mm ($w=10^3$) and 3.67 mm ($w=10^4$)), which demonstrates the effectiveness of PINNs in producing more clinically meaningful deformations. 
\\
\indent  
As shown in Table \ref{TRE values for those patients where DM magnitudes have been corrected}, 22 ($w=10^4$) and 20 ($w=10^3$) out of 77 patients achieved desired smaller \textsf{DMs} in the rigid sub-regions than those in the soft sub-regions with PINNs, while without PINNs for those cases DMs were larger in the rigid sub-regions than those in the soft sub-regions. 
The majority, $68\%$ (15/22) and $80\%$ (16/20) cases, obtained lower TREs than those without PINNs, for $w=10^4$ and $w=10^3$, respectively, where TRE improvements were statistically significant ($p<0.001$) with mean differences
being $2.00$ mm  and $1.88$ mm, respectively. This is consistent with conclusions from previous studies, showing efficacy of imposing distinct material properties within the registration is positively correlated with more accurate registration. \\
\indent Fig. \ref{Root mean squared error on the simulated data} shows results of the second experiment. The \textsf{rmse} values were   $1.90\pm0.52$ mm and $2.11\pm0.63$ mm $(p=0.400)$ for all points,
 $1.94\pm0.59$ mm and $2.19\pm0.75$ mm $(p=0.350)$ for surface points, with PINNs ($w=10^5$) and without PINNs 
 respectively. The enhancements of the PINNs \textit{1)} demonstrate its capability of successfully registering two point sets with lower error values; and \textit{2)} further validate its effectiveness of producing displacement vectors that are more biomechanical compliant, considering that the ground-truth deformations are generated with FEM and thus are implicitly biomechanical encoded. \\
\indent   For the third experiment, compared to that without PINNs, the incorporation of PINNs ($w=10^3$) significantly reduced the average TREs on the test subjects from $6.96\pm1.90$ mm to $6.12\pm1.95$ mm $(p=0.018)$, while Chamfer distances with and without PINNs were $2.48\pm0.33$ mm and $2.54\pm0.37$ mm $(p=0.165)$ on all points
($2.96\pm0.55$ mm and $2.60\pm0.41$ mm $(p<0.001)$ on surface points), respectively. The successful imposition of biomechanical constraints on the test data was further demonstrated by \textit{1)} The ratios of \textsf{DM} between internal points in rigid and soft compartments $\frac{\textsf{DM}_{\text{rigid}}} {\textsf{DM}_{\text{soft}}}$ were $0.89\pm0.11$ and $1.35\pm0.15$ $(p<0.001)$ using registration methods with and without PINNs, respectively; and \textit{2)} 
The loss computed on the test patients using Eq. (\ref{the overall loss function}) was reduced from $20$ to $10^{-14}$ after registration, which demonstrated the network's ability of inferring constraints on unseen subjects. 
\section{Discussions and Conclusions}
\indent  Despite the proposed model's power of regularising biomechanical constraints with predicted transformations and success of reducing registration error and generalising to unseen patients, as we showed in Sect. \ref{Experiments and Results}, this paper needs to be read with several limitations. First, the use of PINNs does not circumvent all limitations of biomechanical modelling shared with other approaches, such as assumptions of potentially subject-specific material properties. However, this opens up new opportunities for solutions to the material property estimation challenge, 
by considering an inverse data-driven discovery problem of PDEs potentially approachable with PINNs \cite{raissi2019physics}.
The second limitation is that our validation is focused on the MRI-TRUS prostate registration, while it is of broad interest to explore the model's effectiveness for wider clinical applications such as accurate and reliable myocardial motion tracking from cardiac cine MRI sequence \cite{QIN2023102682}. The third limitation is that the linear elasticity is assumed, which is useful to demonstrate the efficacy of the methodology but both biomechnical modelling and registration performance may be further improved with more complex modelling with nonlinear materials and geometries in future studies. 
\\
\indent To conclude, in this paper, we have presented a novel biomechanical constraining method using PINNs for non-rigid point set registration. Experimental results on FEM-produced data and clinical MRI-TRUS paired image data, using both patient-specific and multi-patient learning models, demonstrated that the proposed framework is capable of lowering registration errors with presubscribed biomechanical characteristics and generalizability, promising for clinical use and wider research in PINN-based modelling.\\

\section{Acknowledgement}
This work was supported by the Wellcome/EPSRC Centre for Interventional and Surgical Sciences [203145Z/16/Z] and the International Alliance for Cancer Early Detection, an alliance between Cancer Research UK [C28070/A30912; C73666/A31378], Canary Center at Stanford University, the University of Cambridge, OHSU Knight Cancer Institute, University College London and the University of Manchester.

\bibliographystyle{splncs04}
\bibliography{mybibliography}
%






\end{document}